\documentstyle[emulateapj,psfig,natbib,amsfonts,amsmath]{article}
\def\ra#1#2#3{#1$^{\rm h}$#2$^{\rm m}$#3$^{\rm s}$}
\def\dec#1#2#3{$#1^\circ#2'#3''$}
\def\nod{\nodata}

\def\ociw{1}
\def\prince{2}
\def\hubble{3}

\begin{document}

\title{Radio Observations of a Large Sample of Late M, L and T Dwarfs:
The Distribution of Magentic Field Strengths}

\author{
E.~Berger\altaffilmark{\ociw,}\altaffilmark{\prince,}\altaffilmark{\hubble}
}

\altaffiltext{\ociw}{Observatories of the Carnegie Institution
of Washington, 813 Santa Barbara Street, Pasadena, CA 91101}
 
\altaffiltext{\prince}{Princeton University Observatory,
Peyton Hall, Ivy Lane, Princeton, NJ 08544}
 
\altaffiltext{\hubble}{Hubble Fellow}

\begin{abstract} 
We present radio observations of a comprehensive sample of $90$ dwarf
stars and brown dwarfs ranging from spectral type M5 to T8.  We detect
three radio active sources in addition to the six objects previously
detected in quiescence and outburst, leading to an overall detection
rate of about $10\%$ for objects later than M7.  From the properties
of the radio emission we infer magnetic field strengths of $\sim 10^2$
G in quiescence and nearly 1 kG during flares, while the majority of
the non-detected objects have $B\lesssim 50$ G.  Depending on the
configuration and size of the magnetic loops, the surface magnetic
fields may approach 1 kG even in quiescence, at most a factor of few 
smaller than in early-M dwarfs.   With the larger sample
of sources we find continued evidence for (i) a sharp transition
around spectral type M7 from a ratio of radio to X-ray luminosity of
${\rm log}\,(L_R/L_X)\sim -15.5$ to $\gtrsim -12$, (ii) increased
radio activity with later spectral type, in contrast to H$\alpha$ and
X-ray observations, and (iii) an overall drop in the fraction of
active sources from about $30\%$ for M dwarfs to about $5\%$ for L
dwarfs, fully consistent with H$\alpha$ and X-ray observations.  Taken
together, these trends suggest that some late-M and L dwarfs are
capable of generating $0.1-1$ kG magnetic fields, but the overall drop
in the fraction of such objects is likely accompanied by a change in
the structure of the chromospheres and coronae, possibly due to the
increasingly neutral atmospheres and/or a transition to a turbulent
dynamo.  These possibilities can be best addressed through
simultaneous radio, X-ray and H$\alpha$ observations, which can trace
the effect of magnetic fields on the coronae and chromospheres in a
direct, rather than a statistical, manner.  Still, a more extended
radio survey currently holds the best promise for measuring the
magnetic field properties of a large number of dwarf stars.
\end{abstract}
 

\section{Introduction}
\label{sec:intro}

The question of whether and how fully convective low-mass stars and
non-hydrogen burning brown dwarfs generate and dissipate magnetic
fields has important implications for our understanding of their
internal structure and the physical conditions in their atmospheres.
Theoretically, the so-called $\alpha\Omega$ dynamo, which depends on
shearing motions at the radiative-convective transition zone, and
which is used to explain the origin of the solar magnetic field,
cannot operate in fully convective dwarfs.  Instead, the magnetic
dynamo may increasingly depend on turbulent motions associated with
the internal convection itself \citep{rwm+90,ddr93}, but it is not
clear if this mechanism can give rise to a large-scale, long-lived
magnetic field.  Recent magnetohydrodynamic simulations suggest that a
large-scale and stable field may in fact be generated, at least for
parameters roughly appropriate for an M5
dwarf\footnotemark\footnotetext{As noted by \citet{dsb06}, their
fiducial model has properties that are typical of M dwarfs, but with
an artificially high surface temperature due to numerical
limitations.} \citep{dsb06}, contrary to earlier indications
\citep{ddr93}.  It remains to be seen if the same result holds for the
much cooler L and T dwarfs, of which a large fraction lack hydrogen
burning as a source of heat.

In addition, it has been argued that even if strong fields are
generated, their dissipation may be hampered by the increasingly
neutral atmospheres of late-M and L dwarfs \citep{mbs+02}.  If this is
the case, then the suppression of chromospheric and coronal heating
will result in decreased emission in the H$\alpha$ emission line, the
X-rays, and possibly the radio band, despite the presence of magnetic
fields.

Clearly, direct observations which probe the presence, dissipation and
properties of the magnetic fields are required to guide and to test
these theoretical ideas.  Measurements based on Zeeman broadening have
been performed for a handful of active early-M dwarfs (up to M4.5),
indicating field strengths of a few kG with near unity surface
coverage \citep{sl85,jv96}.  Unfortunately, this approach cannot be
used effectively for late-M, L and T dwarfs because the required
atomic lines are weak in the cool atmospheres of these stars and tend
to be blended with molecular features.  It has recently been suggested
that the field strengths can be measured through a secondary
calibration of FeH lines \citep{rb06}, but this approach is yet to be
implemented for any star beyond M4.5.

The presence and dissipation of magnetic fields can be alternatively
traced through activity indicators such as H$\alpha$, X-ray and radio
emission.  The H$\alpha$ and X-ray emission are secondary indicators
since they arise from plasma presumably heated by the dissipation of
magnetic fields, through for example magnetic reconnection.  In the
standard scenario the input of energy drives an outflow of hot plasma
into the corona through evaporation of the underlying chromosphere,
leading in turn to bremsstrahlung X-ray emission and H$\alpha$
emission \citep{neu68,hfs+95,gbs+96}.  The radio emission, on the
other hand, arises from gyroresonance or coherent processes, which
trace the presence and properties of magnetic fields directly.  Thus,
radio observations can be used to infer the field strength of
individual objects directly, whereas H$\alpha$ and X-ray emission
provide a useful statistical measure and insight into the influence of
magnetic fields on the outer layers of dwarf stars.

Observationally, the lack of significant change in the measured level
of H$\alpha$ and X-ray activity with the onset of full convection at
about spectral type M3, suggests that at least in the early-M dwarfs
the putative turbulent or distributed dynamo can operate efficiently.
It is also possible that the magnetic field itself acts to reduce the
mass at which the transition to fully convective structure takes place
\citep{mm01}.  However, beyond spectral type M7 there is a precipitous
drop in H$\alpha$ and X-ray persistent activity, and only a few
percent of the objects exhibit flares (e.g.,
\citealt{rkg+99,gmr+00,rbm+00,lkc+03,whw+04}).  Furthermore, unlike in
the early-M dwarfs \citep{rgv85,fgs+93,mbs+02,pmm+03}, many late-type
rapidly rotating dwarfs exhibit little or no discernible activity in
these bands \citep{bm95,mb03}.  These patterns are consistent with
either a decrease in the generation or dissipation of the magnetic
fields, or both.

On the other hand, radio emission has been detected from several
late-M and L dwarfs \citep{bbb+01,ber02,brr+05,bp05} suggesting that at 
least some of these objects are capable of generating and dissipating
magnetic fields.  Surprisingly, the ratio of radio to X-ray luminosity
in the detected objects exceeds by several orders of magnitude the
value measured for early M dwarfs and a variety of other stars
(including the Sun; \citealt{gb93,bg94}), and there is no obvious
correlation with H$\alpha$ emission.  Thus, radio observations present
a powerful, and perhaps unique approach for inferring the magnetic
field strength of late-type stars and brown dwarfs.

Here we exploit this approach and continue our investigation of radio
emission from late-M, L and T dwarfs by expanding the observed sample
by about a factor of three (to $90$ sources).  With this extended
sample we find continued evidence for a sharp transition in the ratio
of radio to X-ray luminosity at spectral type M7, as well as an
increased level of activity with later spectral type.  We show,
however, that as in the case of H$\alpha$ and X-ray observations, the
fraction of objects producing radio emission drops from about $30\%$
in the M dwarfs to only $\sim 5\%$ in the L dwarfs.  Most importantly,
we present for the first time estimates of the magnetic field strength
of a large sample of late-M and L dwarfs, and show that for the active
sources there is at most a modest drop in the field strength from
early-M to early-L dwarfs.

\section{Observations}
\label{sec:obs}

\begin{figure}
\epsscale{0.9}
\centerline{\psfig{file=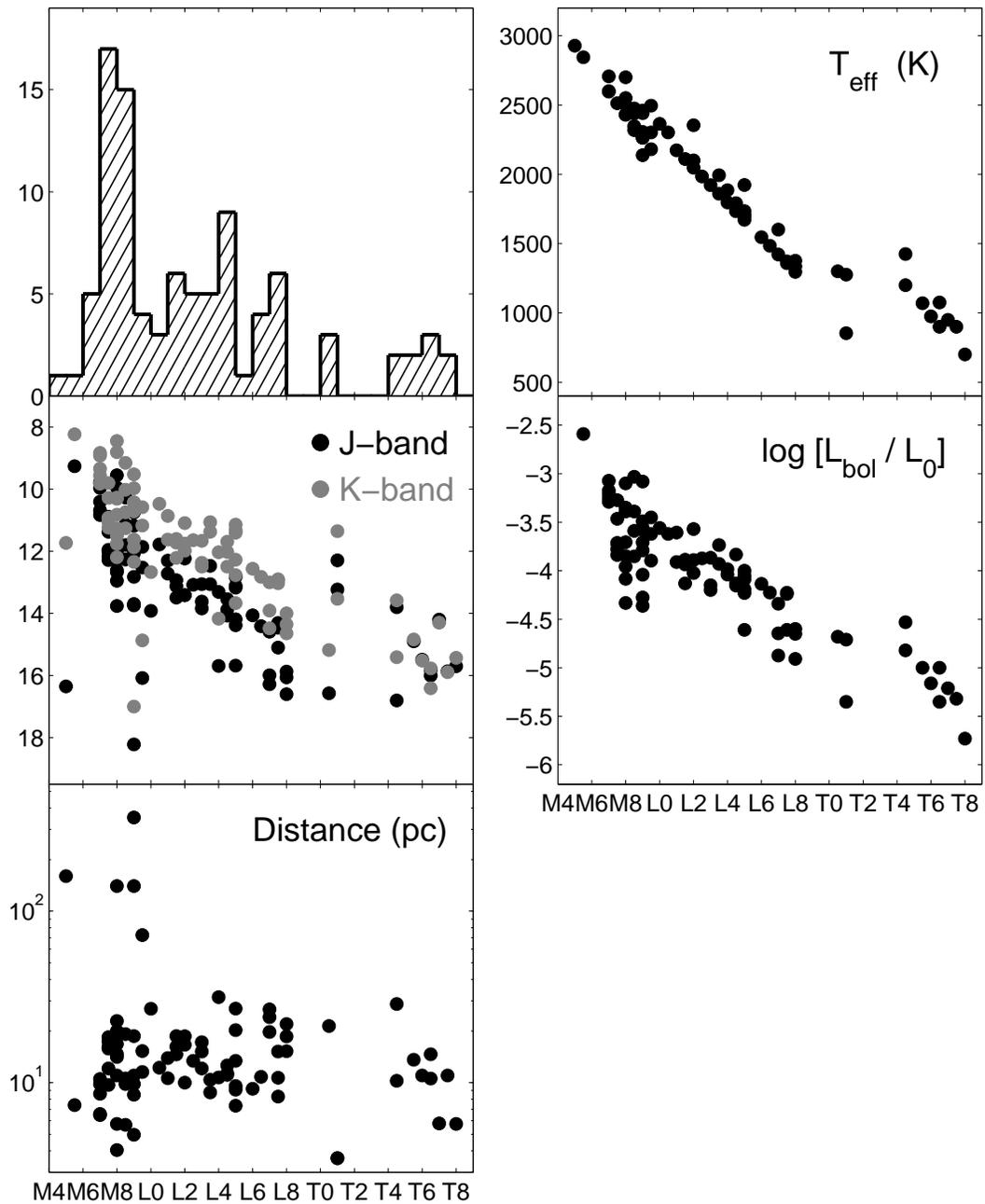,width=5.5in}}
\caption{General properties of the survey sources, including spectral
type, near-IR magnitudes, distances, effective temperatures, and
bolometric luminosities.
\label{fig:prop}}
\end{figure}

We observed a sample of $21$ late-M, L and T dwarfs with the Very
Large Array (VLA\footnotemark\footnotetext{The VLA is operated by the
National Radio Astronomy Observatory, a facility of the National
Science Foundation operated under cooperative agreement by Associated
Universities, Inc.}) at 8.46 GHz using the standard continuum mode
with $2\times 50$ MHz contiguous bands at each frequency.  The flux
density scale was determined using the standard extragalactic
calibrator sources 3C 48 (J0137+331), 3C 147 (J0542+498) and 3C 286
(J1331+305), while the phase was monitored using calibrators located
within $10^\circ$ of the targets sources.  The data were reduced and
analyzed using the Astronomical Image Processing System.  In addition
to our observations we obtained and reduced all publicly available
observations of late-M, L and T dwarfs from the VLA archive, and
collected all measurements published in the literature.  This resulted
in a total 88 objects ranging from M7 to T8, as well a single M5 dwarf
and a single M5.5 dwarf.  A summary of all observations and the
relevant source properties are given in Table~\ref{tab:obs} and
Figure~\ref{fig:prop}.

For the detected objects we searched for variability (flares) using
the following method.  We removed all the bright field sources using
the AIPS/IMAGR routine to CLEAN the region around each source (with
the exception of the target source), and the AIPS/UVSUB routine to
subtract the resulting source models from the visibility data.  We
then plotted the real part of the complex visibilities at the position
of the science target as a function of time using the AIPS/UVPLT
routine.  The subtraction of field sources is necessary since their
sidelobes and the change in the shape of the synthesized beam during
the observation result in flux variations over the map, which may
contaminate any real variability or generate false variability.

\section{Properties of the Radio Emission} 
\label{sec:rad}

Quiescent radio emission has been previously detected from six
late-type dwarfs ranging from spectral type M7 to L3.5
\citep{bbb+01,ber02,brr+05,bp05,ohb+06}.  Four of these objects also
produced short-lived, highly-polarized flares with a typical timescale
of $\sim 10$ min and a flux increase compared to the quiescent level
of at least a factor of few.  In addition, the L3.5 dwarf 2MASS
J00361617+1821104 was shown to exhibit a periodicity of 3 hr in its
quiescent radio emission, whose origin is not fully understood, but
may arise from a closely-orbiting companion \citep{brr+05}.

In the extended sample we detect radio emission from three additional
dwarf stars: LHS 1070 (M5.5), LSR J1835+3259 (M8.5), and 2MASS
J$05233822-1403022$ (L2.5) with fluxes of $161\pm 15$ $\mu$Jy, $525\pm
15$ $\mu$Jy and $231\pm 14$ $\mu$Jy, respectively.  LHS 1070 has been
detected on two separate occasions with fluxes of $153\pm 23$ and
$167\pm 20$ $\mu$Jy, respectively, consistent with non-variable
quiescent emission.  Similarly, we do not find evidence for
variability within a single observation for any of the three sources,
but we note that the observations are relatively short, $\approx
90-220$ min.  The fractional circular polarization for the three
objects is $f_c<30\%$ (LHS 1070), $f_c<9\%$ (LSR J1835+3259), and
$f_c=19\pm 6\%$ (2M\,$0523-14$), similar to the level of circular
polarization in the quiescent emission from previous objects.

Finally, we note that weak radio emission is detected in near
positional coincidence ($\lesssim 5''$) with two other sources, 2MASS
J$01483864-3024396$ (M7.5) and 2MASS J$04234858-0414035$ (L7.5).
However, we do not believe that these are genuine detections for two
reasons.  First, the offset for 2M\,$0423-04$ does not coincide with
the proper motion measured by \citet{vhl+04}; for 2M J$0148-30$ the
proper motion is not known but the offset is $3.3\pm 0.4''$ with a
position angle of about $75^\circ$.  Second, in both cases the
emission is strongly detected in only one of the two intermediate
frequency (IF) channels of the VLA, but is consistent with zero in the
other IF channel.  This suggests that the radio emission is most
likely spurious and may be due to a low level of interference.

\section{Physical Properties: Magnetic Field Strength, Source Size,
and Density}
\label{sec:phys}

Using the observed fluxes and fractional circular polarization we now
estimate the magnetic field strengths for the detected sources.  We
first provide a rough estimate of the brightness temperature as a way
to assess the origin of the radio emission (coherent
vs.~gyrosynchrotron): $T_b=2\times 10^9\,F_{\rm \nu,mJy}\,\nu_{\rm
GHz}^{-2}\,d_{\rm pc}^2\,(R/R_{\rm J})^{-2}\,{\rm K}$, where
$R_J\approx R_s\approx 7\times 10^9$ cm is Jupiter's radius and
roughly the source radius, and $R\sim (1-2)R_J$ \citep{lpl+00} is the
size of the emitting region if the covering fraction is of order
unity.  For our detected sources we find $T_b\approx 10^8-10^9$ K.
Conversely, the inverse Compton limit of $T_b\lesssim 10^{12}$ K for
gyrosynchrotron emission defines a minimum size for the emitting
region of $R\approx 0.035R_s \approx 2.5\times 10^8$ cm, or about
$0.03\%$ of the surface area at the height of the corona.  If the
emitting region is in fact more compact than this size, then the
emission is most likely due to coherent emission processes such as
plasma radiation or electron cyclotron maser.  Since the latter
emission mechanisms are typically responsible for short-lived flares
we proceed with the reasonable assumption that the observed emission
is due to gyrosynchrotron radiation.

In this context we follow the typical assumption that the emitting
electrons obey a power law distribution, $N(\gamma)\propto
\gamma^{-p}$ above a cutoff Lorentz factor, $\gamma_{\rm m}$.  The
value of the power law index is typically $p\sim 3$ for M dwarfs, and
was found to be bout 2.7 for the L3.5 dwarf 2M\,0036+18
\citep{brr+05}.  We adopt the standard value of $p=3$ here.  The peak
frequency, flux, and degree of circular polarization of the detected
radio emission are directly related to the density of emitting
electrons ($n_e$), the size of the emission region ($R$), and the
magnetic field strength ($B$).  Following the formulation of
\citet{gud02} we find that the peak frequency is given by
\begin{equation}
\nu_m\approx 1.66\times 10^4\,n_e^{0.23}\,R^{0.23}\,B^{0.77}\,\, 
{\rm Hz},
\end{equation}
the flux density is given by
\begin{equation}
F_{\nu,m}\approx 1.54\times 10^{-4}\,B^{-0.76}\,R^2\,d^{-2}\,
\nu_m^{2.76}\,\, {\rm \mu Jy},
\end{equation}
the fraction of circular polarization is given by
\begin{equation}
f_c\approx 2.85\times 10^3\,B^{0.51}\,\nu_m^{-0.51},
\end{equation}
and we assume an average angle of $\pi/3$ between the magnetic field
and the line of sight.

\begin{figure}
\epsscale{1}
\centerline{\psfig{file=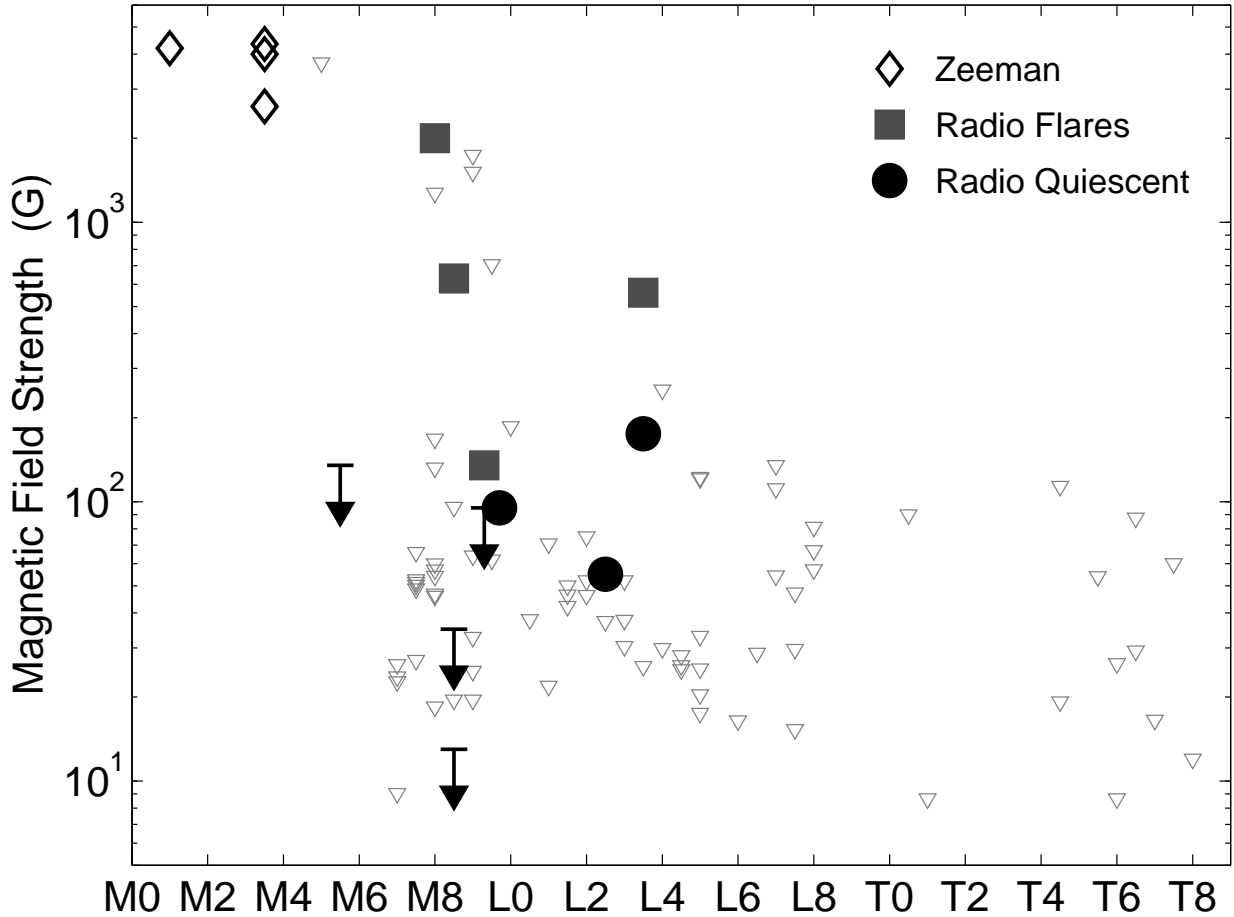,width=6.5in}}
\caption{Magnetic field strengths and upper limits as inferred from 
the radio observations (quiescent: circles; flares: squares).  Also 
shown are the values for early M dwarfs measured from Zeeman line 
broadening \citep{sl85,jv96}.  We note that the surface field
strength in the radio active dwarfs may up to an order of magnitude 
larger depending on the structure of the field and the height of the 
radio emitting region.
\label{fig:mag}}
\end{figure}

Using $\nu_m=8.5$ GHz as an estimate of the spectral peak, and the
measured circular polarization fractions we find that $B\approx 55\pm
17$ G for 2M\,$0523-14$, $B<135$ G for LHS 1070, and $B<13$ G for LSR
J1835+3259.  For 2M\,$0523-14$ we further find a source size $R\approx
4.6\times 10^9\,{\rm cm}\approx 0.7R_J$, and $n_e\approx 2.2\times
10^9$ cm$^{-3}$.  The former indicates that at the height of the
emission region (presumably the corona) the field covers $\sim 5-10\%$
of the surface area.  For the two other sources we find $R\lesssim
3.0\times 10^9$ cm and $n_e\gtrsim 3.3\times 10^9$ cm$^{-3}$ (LHS
1070) and $R\lesssim 1.7\times 10^9$ cm and $n_e\gtrsim 5.8\times
10^9$ cm$^{-3}$ (LSR J1835+3259).  These values, along with those
inferred for the six dwarf stars detected previously, are summarized
in Table~\ref{tab:phys} and Figure~\ref{fig:mag}.  We note that the
inferred sizes are consistent with our inference based on the 
brightness temperature argument that the radio emission is due to 
gyrosynchrotron radiation,

For the non-detected sources we find a rough limit on the magnetic
field strength by assuming typical values of the source size, $0.5
R_J$, and electron density, $n_e=10^9$ cm$^{-3}$, as inferred from the
detected sources.  The resulting upper limit on the magnetic field
strength is thus $B<9\,F_{\rm \nu,mJy}^{0.73}\,d_{\rm pc}^{1.46}$ G,
which at our typical sensitivity threshold corresponds to $B<10^2$ G
for $d\lesssim 25$ pc; for objects within 10 pc, the limit is
$B\lesssim 30$ G (Figure~\ref{fig:mag}).

Thus, we find from both the detected and non-detected objects that the
typical quiescent magnetic fields in late-M, L and T dwarfs are of
the order of $\lesssim 10^2$ G, but may approach $\sim 1$ kG during
flares.  We note that these field strengths are relevant at the
location where the radio emission is produced, possibly $\sim
1-2\,R_s$ above the stellar surface.  Thus, depending on the exact
field configuration, the surface magnetic field may be an order of
magnitude larger, or nearly 1 kG even in quiescence.

The inferred field strengths can be compared to those of a few early M
dwarfs (EV Lac, AD Leo, AU Mic, Gl 729) for which values of about 4 kG
with a surface coverage approaching unity have been inferred from
Zeeman line broadening \citep{sl85,jv96}.  Somewhat weaker fields,
$\sim 1$ kG, have been estimated for young accreting brown dwarfs
based on evidence for magnetic funneling from variations in the
H$\alpha$ line \citep{sj06}.  Thus, it appears that the magnetic field
strengths in field late-M, L and T dwarfs may be a factor of few
smaller than in early-M dwarfs, and possibly with a smaller surface
coverage, but these fully convective stars and brown dwarfs are
clearly capable of generating large-scale, long-lived fields.

\section{Trends and Implications}
\label{sec:trend}

With the large sample of $90$ M5--T8 dwarfs presented in this paper
and compiled from previous work, we can begin to address trends in the
radio emission, and hence magnetic properties of dwarf stars.  Three
interesting possibilities have been suggested previously based on a
smaller sample of sources \citep{ber02}.  First, the ratio of radio to
X-ray luminosity of late-M and L dwarfs appears to be orders of
magnitude larger than that of other stars, including M dwarfs earlier
than spectral type M7.  Second, the level of radio activity appears to
increase with later spectral type.  Finally, there may be a
correlation between the strength of the radio activity and rotation
velocity, such that rapid rotators exhibit stronger radio activity.

We first investigate any trends in the strength of the radio activity
as a function of spectral type.  A proper comparison requires
normalization by the bolometric luminosity of each source.  These are
available in the literature for about half of the survey sources,
including all of the T dwarf.  For the rest we use published
bolometric correction factors.  For the L dwarfs $BC_K=3.42+0.075
({\rm SP}-4)$ for L0--L4 and $BC_K=3.42-0.075({\rm SP}-4)$ for L5--L9,
with ${\rm SP}=0$ for L0 \citep{dhv+02,nty04}.  For the M dwarfs we
use the mean of $BC_J=2.43+ 0.0895{\rm SP}$ and $BC_K=1.53+0.148{\rm
SP}-0.0105{\rm SP}^2$, with ${\rm SP}=0$ for M0 \citep{wgm99}.

\begin{figure}
\epsscale{1}
\centerline{\psfig{file=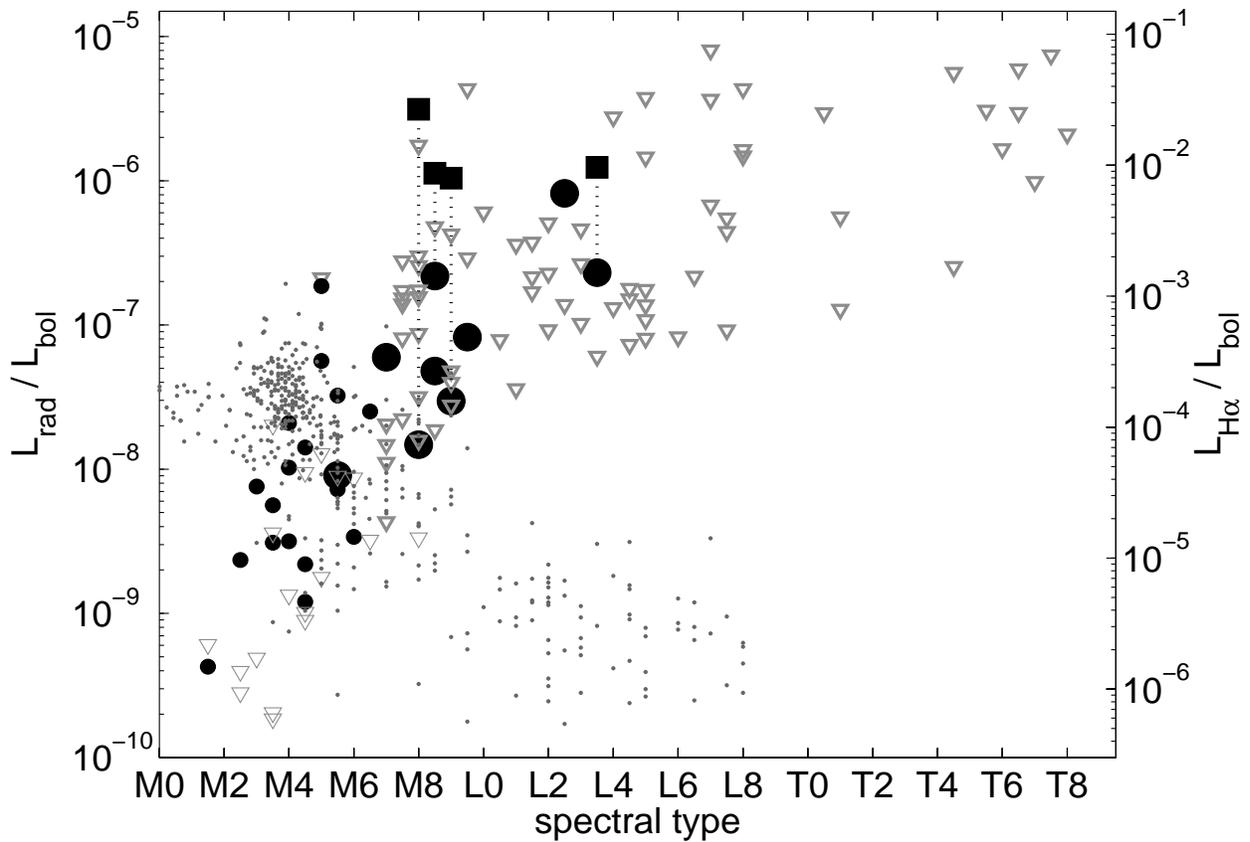,width=6.5in}}
\caption{Radio and H$\alpha$ activity as a function of spectral type.
Shown are flares (squares), quiescent emission (circles), and upper
limits (triangles) in the radio; the H$\alpha$ observations are shown
as gray dots.  Note that the scale on the ordinate is different for
the radio and H$\alpha$ observations.  The trend of increased radio
activity with later spectral type is evident, as is the overall drop
in the fraction of detected objects.  Unlike the radio emission, the
H$\alpha$ emission drops beyond spectral type of about M7.
\label{fig:activity}}
\end{figure}

With these values, we plot the ratio of radio to bolometric
luminosity, $L_{\rm rad}/L_{\rm bol}$, as a function of spectral type
in Figure~\ref{fig:activity}.  Two trends are clear from this figure.
First, the fraction of detected objects drops considerably as a
function of spectral type, from about $30\%$ for the M dwarfs, to
about $4\%$ for the L dwarfs.  If we consider only non-detections that
are lower than the level of activity in the detected objects, these
fractions are roughly $45\%$ and $7\%$, respectively.  Second, and
perhaps more important, the level of radio activity tends to increase
with later spectral types, with the active L dwarfs exhibiting a level
of activity at least an order of magnitude larger than the mid-M
dwarfs.  Whether this trend continues to late-L and T spectral types
remains unclear due to the relatively small number of observed objects
and the observed decrease in the fraction of active sources.  Still,
the increased level of activity supports our inference that the
magnetic field strengths are note significantly lower in late-M and L
dwarfs.

On the other hand, with the larger sample we do not find clear
evidence for a correlation between the level of activity and rotation
velocity.  Specifically, for the seven detected sources with a
spectral type later than M7 and a measured rotation velocity, the
average value of $L_{\rm R}/L_{\rm bol}$ is roughly the same for
$v{\rm sin}i<20$ km s$^{-1}$ and $>20$ km s$^{-1}$.  A more careful
analysis of this possible trend requires rotation velocity
measurements for a significantly larger sample.  As can be seen from
Table~\ref{tab:obs}, less than a quarter of the observed objects have
measured velocities.

\begin{figure}
\epsscale{1}
\centerline{\psfig{file=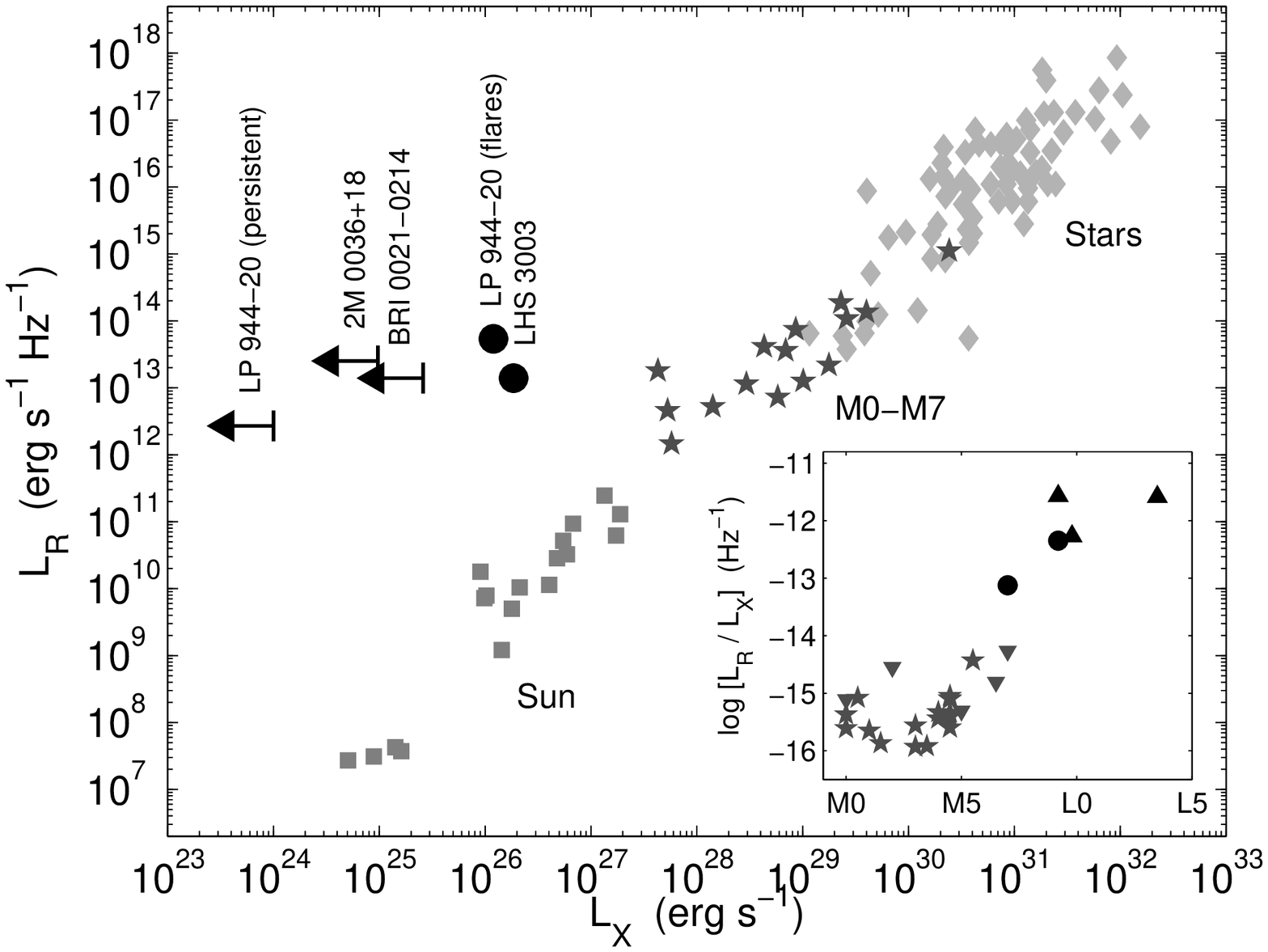,width=6.5in}}
\caption{Radio vs.~X-ray luminosity for stars exhibiting coronal
activity.  Data for late-M and L dwarfs are from \citet{rbm+00},
\citet{bbb+01}, \citet{ber02}, \citet{brr+05}, \citet{bp05}, while
other data are taken from \citet{gud02} and references therein.  Data
points for the Sun include impulsive and gradual flares, as well as
microflares.  The strong correlation between $L_{R}$ and $L_X$ is
evident, but begins to break down around spectral type M7 (see inset).
The late-M and L dwarfs clearly violate the correlation and are
over-luminous in the radio.
\label{fig:gb}}
\end{figure}

Finally, we find continued evidence that the correlation between radio
and X-ray luminosity, which is roughly constant at a value of ${\rm
log}\,(L_R/L_X)\sim -15.5$ for a wide range of stars
\citep{gb93,bg94}, breaks down at spectral type of about M7
(Figure~\ref{fig:gb}).  Possible explanations for the breakdown in
this correlation have been discussed previously
\citep{ber02,brr+05,ohb+06}, and focus on efficient trapping of the
relativistic electrons, thereby reducing the efficiency of coronal and
chromospheric heating, or inefficient production of X-rays due to a
reduction in the number of free ions in the cool atmospheres of these
dwarf stars.  From a purely observational standpoint, the data
indicate that the decrease in X-ray luminosity by three orders of
magnitude occurs over a narrow range in spectral type at about M7 (or
$T_{\rm eff}\approx 2500-2700$ K), roughly the same spectral type
where a transition to predominantly neutral atmospheres occurs
\citep{mbs+02}.  This is also the same location at which a significant
drop in H$\alpha$ activity occurs, from an average value of ${\rm
log}\,(L_{\rm H\alpha}/L_{\rm bol})\approx -3.6$ earlier than M5 to
$\approx -4.3$ later than M7 \citep{whw+04}.  The fact that the
magnetic field strengths and radio activity do not decrease
significantly indicates that the drop in X-ray and H$\alpha$ activity
is related instead to the structure of the atmosphere, chromosphere
and corona.

\section{Conclusions and Future Directions}
\label{sec:conc}

We presented radio observations of 88 dwarf stars and brown dwarfs in
the range M7--T8, along with a single M5 and a single M5.5 dwarf.
This sample is nearly comparable to the number of objects with
H$\alpha$ measurements, but is significantly larger than the number of
objects observed in the X-rays.  The drop in the fraction of active
sources at the M/L transition, which is observed in H$\alpha$, is also
apparent in the radio observations.  However, the strength of the
activity in the detected objects is in fact higher for the L dwarfs,
both in quiescence and during flares.  As we have shown in
\S\ref{sec:phys}, this is probably because the magnetic field
strengths remain roughly unchanged despite the drop in effective
temperature and luminosity with later spectral type.  The implication
is that the magnetic dynamo process may be similar in early-M and L
dwarfs, and is not strongly dependent on luminosity or temperature.

We also stress that while H$\alpha$ observations are relatively simple
to carry out, it is not trivial to translate the observed emission to
an estimate of the magnetic field strength.  The advantage of radio
observations is that they directly trace the strength of the field,
and provide in addition information on the physical properties of the
emission region, such as its size and density.  We therefore suggest
that a three-pronged approach is required for continued progress in
our understanding of magnetic fields in dwarf stars and brown dwarfs.

First, the survey for radio emission from these objects should be
expanded to all of the $\sim 500$ currently known L and T dwarfs, as
well as a large sample of objects in the spectral range M5--M9, which
covers the breakdown in the radio/X-ray correlation.  Such a survey
can be carried out efficiently with the EVLA, which is scheduled to
come on-line in the next few years and which will deliver a nearly
order of magnitude increase in sensitivity.  This is essential since
the majority of the current detections are near the threshold of the
VLA.  We estimate that about $500-1,000$ hr of observing time would be
required for such an undertaking, delivering a sensitivity of $L_{\rm
rad}/L_{\rm bol}\sim {\rm few}\times 10^{-9}$ at L0 and $\sim 10^{-8}$
at T0.

Second, the objects that have been detected so far, and will be
detected with the expanded survey, should be observed over a wide
frequency range to better characterize the shape of their spectrum and
the frequency dependence of their polarization.  This is essential for
deriving magnetic field strengths to better accuracy than is possible
with the current single-frequency observations.  We expect that at the
current signal-to-noise level, the magnetic field strengths can be
inferred to an accuracy of $\sim 20-30\%$, which is comparable to, or
better than, measurements made from Zeeman broadening for early-M
dwarfs.  In addition, long timescale, high signal-to-noise
observations can be used to check for possible periodicity in the
radio emission, as has been uncovered for the L3.5 dwarf 2M\,0036+18
\citep{brr+05}.  If the periodicity is in fact related to a close-in
companion, which excites the dynamo by tidal or magnetic interactions,
this may be a ubiquitous feature of the active objects.

Finally, these same objects should be observed simultaneously in the
radio, X-rays, and H$\alpha$ in order to directly measure the
correlation, or lack thereof, between these activity indicators.  This
is necessary in order to trace the origin of the shift in the
radio/X-ray correlation, and in order to trace the evolution of flares
as the release of magnetic stresses, evident in the radio, heats up
the corona (X-rays) and chromosphere (H$\alpha$).  While the current
observations allow us to address in a statistical manner the overall
trends observed with each technique, only simultaneous observations
can provide insight into the production and evolution of flares.  As
the most catastrophic events in the atmospheres of dwarf stars, such
events will undoubtedly shed light on the structure of the fields,
their strengths, and the details of the energy dissipation process,
all of which will provide observational constraints on the dynamo
mechanism in dwarf stars.

\acknowledgements 
Research has benefitted from the M, L, and T dwarf compendium housed
at DwarfArchives.org and maintained by Chris Gelino, Davy Kirkpatrick,
and Adam Burgasser.  This work has made use of the SIMBAD database,
operated at CDS, Strasbourg, France.  E.B.~is supported is
supported by NASA through Hubble Fellowship grant HST-01171.01 awarded
by the Space Telescope Science Institute, which is operated by AURA,
Inc., for NASA under contract NAS 5-26555.


\begin{thebibliography}{}

\bibitem[\protect\citeauthoryear{{Audard} et~al.}{{Audard}
  et~al.}{2005}]{abb+05}
{Audard}, M., {Brown}, A., {Briggs}, K.~R., {G{\"u}del}, M., {Telleschi}, A.,
  \& {Gizis}, J.~E. 2005, \apjl, 625, L63

\bibitem[\protect\citeauthoryear{{Bailer-Jones}}{{Bailer-Jones}}{2004}]{bai04}
{Bailer-Jones}, C.~A.~L. 2004, \aap, 419, 703

\bibitem[\protect\citeauthoryear{{Basri} \& {Marcy}}{{Basri} \&
  {Marcy}}{1995}]{bm95}
{Basri}, G.,  \& {Marcy}, G.~W. 1995, AJ, 109, 762

\bibitem[\protect\citeauthoryear{{Benz} \& {Guedel}}{{Benz} \&
  {Guedel}}{1994}]{bg94}
{Benz}, A.~O.,  \& {Guedel}, M., 285, 621

\bibitem[\protect\citeauthoryear{{Berger}}{{Berger}}{2002}]{ber02}
{Berger}, E. 2002, \apj, 572, 503

\bibitem[\protect\citeauthoryear{{Berger} et~al.}{{Berger}
  et~al.}{2001}]{bbb+01}
{Berger}, E., et~al. 2001, \nat, 410, 338

\bibitem[\protect\citeauthoryear{{Berger} et~al.}{{Berger}
  et~al.}{2005}]{brr+05}
{Berger}, E., et~al. 2005, \apj, 627, 960

\bibitem[\protect\citeauthoryear{{Bouy} et~al.}{{Bouy} et~al.}{2003}]{bbm+03}
{Bouy}, H., {Brandner}, W., {Mart{\'{\i}}n}, E.~L., {Delfosse}, X., {Allard},
  F.,  \& {Basri}, G. 2003, \aj, 126, 1526

\bibitem[\protect\citeauthoryear{{Bouy} et~al.}{{Bouy} et~al.}{2004}]{bdk+04}
{Bouy}, H., et~al. 2004, \aap, 423, 341

\bibitem[\protect\citeauthoryear{{Burgasser} et~al.}{{Burgasser}
  et~al.}{1999}]{bkb+99}
{Burgasser}, A.~J., et~al. 1999, \apjl, 522, L65

\bibitem[\protect\citeauthoryear{{Burgasser} et~al.}{{Burgasser}
  et~al.}{2003a}]{bkl+03}
{Burgasser}, A.~J., {Kirkpatrick}, J.~D., {Liebert}, J.,  \& {Burrows}, A.
  2003a, \apj, 594, 510

\bibitem[\protect\citeauthoryear{{Burgasser} et~al.}{{Burgasser}
  et~al.}{2003b}]{bkm+03}
{Burgasser}, A.~J., {Kirkpatrick}, J.~D., {McElwain}, M.~W., {Cutri}, R.~M.,
  {Burgasser}, A.~J.,  \& {Skrutskie}, M.~F. 2003b, \aj, 125, 850

\bibitem[\protect\citeauthoryear{{Burgasser} et~al.}{{Burgasser}
  et~al.}{2003c}]{bkr+03}
{Burgasser}, A.~J., {Kirkpatrick}, J.~D., {Reid}, I.~N., {Brown}, M.~E.,
  {Miskey}, C.~L.,  \& {Gizis}, J.~E. 2003c, \apj, 586, 512

\bibitem[\protect\citeauthoryear{{Burgasser} \& {Putman}}{{Burgasser} \&
  {Putman}}{2005}]{bp05}
{Burgasser}, A.~J.,  \& {Putman}, M.~E. 2005, \apj, 626, 486

\bibitem[\protect\citeauthoryear{{Burgasser} et~al.}{{Burgasser}
  et~al.}{2005}]{brl+05}
{Burgasser}, A.~J., {Reid}, I.~N., {Leggett}, S.~K., {Kirkpatrick}, J.~D.,
  {Liebert}, J.,  \& {Burrows}, A. 2005, \apjl, 634, L177

\bibitem[\protect\citeauthoryear{{Burgasser} et~al.}{{Burgasser}
  et~al.}{2000}]{bwk+00}
{Burgasser}, A.~J., et~al. 2000, \aj, 120, 1100

\bibitem[\protect\citeauthoryear{{Costa} et~al.}{{Costa} et~al.}{2005}]{cmj+05}
{Costa}, E., {M{\'e}ndez}, R.~A., {Jao}, W.-C., {Henry}, T.~J., {Subasavage},
  J.~P., {Brown}, M.~A., {Ianna}, P.~A.,  \& {Bartlett}, J. 2005, \aj, 130, 337

\bibitem[\protect\citeauthoryear{{Cruz} \& {Reid}}{{Cruz} \&
  {Reid}}{2002}]{cr02}
{Cruz}, K.~L.,  \& {Reid}, I.~N. 2002, \aj, 123, 2828

\bibitem[\protect\citeauthoryear{{Cruz} et~al.}{{Cruz} et~al.}{2003}]{crl+03}
{Cruz}, K.~L., {Reid}, I.~N., {Liebert}, J., {Kirkpatrick}, J.~D.,  \&
  {Lowrance}, P.~J. 2003, \aj, 126, 2421

\bibitem[\protect\citeauthoryear{{Dahn} et~al.}{{Dahn} et~al.}{2002}]{dhv+02}
{Dahn}, C.~C., et~al. 2002, \aj, 124, 1170

\bibitem[\protect\citeauthoryear{{Deacon} \& {Hambly}}{{Deacon} \&
  {Hambly}}{2001}]{dh01}
{Deacon}, N.~R.,  \& {Hambly}, N.~C. 2001, \aap, 380, 148

\bibitem[\protect\citeauthoryear{{Delfosse} et~al.}{{Delfosse}
  et~al.}{1999}]{dtf+99}
{Delfosse}, X., {Tinney}, C.~G., {Forveille}, T., {Epchtein}, N.,
  {Borsenberger}, J., {Fouqu{\'e}}, P., {Kimeswenger}, S.,  \& {Tiph{\`e}ne},
  D. 1999, \aaps, 135, 41

\bibitem[\protect\citeauthoryear{{Dobler}, {Stix}, \& {Brandenburg}}{{Dobler}
  et~al.}{2006}]{dsb06}
{Dobler}, W., {Stix}, M.,  \& {Brandenburg}, A. 2006, \apj, 638, 336

\bibitem[\protect\citeauthoryear{{Durney}, {De Young}, \& {Roxburgh}}{{Durney}
  et~al.}{1993}]{ddr93}
{Durney}, B.~R., {De Young}, D.~S.,  \& {Roxburgh}, I.~W., 145, 207

\bibitem[\protect\citeauthoryear{{Fleming} et~al.}{{Fleming}
  et~al.}{1993}]{fgs+93}
{Fleming}, T.~A., {Giampapa}, M.~S., {Schmitt}, J.~H.~M.~M.,  \& {Bookbinder},
  J.~A. 1993, \apj, 410, 387

\bibitem[\protect\citeauthoryear{{Fuhrmeister} \& {Schmitt}}{{Fuhrmeister} \&
  {Schmitt}}{2004}]{fs04}
{Fuhrmeister}, B.,  \& {Schmitt}, J.~H.~M.~M. 2004, \aap, 420, 1079

\bibitem[\protect\citeauthoryear{{Geballe} et~al.}{{Geballe}
  et~al.}{2002}]{gkl+02}
{Geballe}, T.~R., et~al. 2002, \apj, 564, 466

\bibitem[\protect\citeauthoryear{{Gizis}}{{Gizis}}{2002}]{giz02}
{Gizis}, J.~E. 2002, \apj, 575, 484

\bibitem[\protect\citeauthoryear{{Gizis} et~al.}{{Gizis} et~al.}{2000}]{gmr+00}
{Gizis}, J.~E., {Monet}, D.~G., {Reid}, I.~N., {Kirkpatrick}, J.~D., {Liebert},
  J.,  \& {Williams}, R.~J. 2000, \aj, 120, 1085

\bibitem[\protect\citeauthoryear{{Gizis} \& {Reid}}{{Gizis} \&
  {Reid}}{2006}]{gr06}
{Gizis}, J.~E.,  \& {Reid}, I.~N. 2006, \aj, 131, 638

\bibitem[\protect\citeauthoryear{{Golimowski} et~al.}{{Golimowski}
  et~al.}{2004}]{glm+04}
{Golimowski}, D.~A., et~al. 2004, \aj, 127, 3516

\bibitem[\protect\citeauthoryear{{Graham} et~al.}{{Graham}
  et~al.}{1992}]{gmg+92}
{Graham}, J.~R., {Matthews}, K., {Greenstein}, J.~L., {Neugebauer}, G.,
  {Tinney}, C.~G.,  \& {Persson}, S.~E. 1992, \aj, 104, 2016

\bibitem[\protect\citeauthoryear{{G{\"u}del}}{{G{\"u}del}}{2002}]{gud02}
{G{\"u}del}, M. 2002, \araa, 40, 217

\bibitem[\protect\citeauthoryear{{Guedel} \& {Benz}}{{Guedel} \&
  {Benz}}{1993}]{gb93}
{Guedel}, M.,  \& {Benz}, A.~O. 1993, ApJ, 405, L63

\bibitem[\protect\citeauthoryear{{Guedel} et~al.}{{Guedel}
  et~al.}{1996}]{gbs+96}
{Guedel}, M., {Benz}, A.~O., {Schmitt}, J.~H.~M.~M.,  \& {Skinner}, S.~L. 1996,
  \apj, 471, 1002

\bibitem[\protect\citeauthoryear{{Hawley} et~al.}{{Hawley}
  et~al.}{2002}]{hck+02}
{Hawley}, S.~L., et~al. 2002, \aj, 123, 3409

\bibitem[\protect\citeauthoryear{{Hawley} et~al.}{{Hawley}
  et~al.}{1995}]{hfs+95}
{Hawley}, S.~L., et~al. 1995, \apj, 453, 464

\bibitem[\protect\citeauthoryear{{Johns-Krull} \& {Valenti}}{{Johns-Krull} \&
  {Valenti}}{1996}]{jv96}
{Johns-Krull}, C.~M.,  \& {Valenti}, J.~A. 1996, \apjl, 459, L95

\bibitem[\protect\citeauthoryear{{Kendall} et~al.}{{Kendall}
  et~al.}{2004}]{kdm+04}
{Kendall}, T.~R., {Delfosse}, X., {Mart{\'{\i}}n}, E.~L.,  \& {Forveille}, T.
  2004, \aap, 416, L17

\bibitem[\protect\citeauthoryear{{Kendall} et~al.}{{Kendall}
  et~al.}{2003}]{kma+03}
{Kendall}, T.~R., {Mauron}, N., {Azzopardi}, M.,  \& {Gigoyan}, K. 2003, \aap,
  403, 929

\bibitem[\protect\citeauthoryear{{Kirkpatrick}}{{Kirkpatrick}}{2005}]{kir05}
{Kirkpatrick}, J.~D. 2005, \araa, 43, 195

\bibitem[\protect\citeauthoryear{{Kirkpatrick} et~al.}{{Kirkpatrick}
  et~al.}{2001}]{kdm+01}
{Kirkpatrick}, J.~D., {Dahn}, C.~C., {Monet}, D.~G., {Reid}, I.~N., {Gizis},
  J.~E., {Liebert}, J.,  \& {Burgasser}, A.~J. 2001, \aj, 121, 3235

\bibitem[\protect\citeauthoryear{{Kirkpatrick}, {Henry}, \&
  {Liebert}}{{Kirkpatrick} et~al.}{1993}]{khl93}
{Kirkpatrick}, J.~D., {Henry}, T.~J.,  \& {Liebert}, J. 1993, \apj, 406, 701

\bibitem[\protect\citeauthoryear{{Kirkpatrick} et~al.}{{Kirkpatrick}
  et~al.}{2000}]{krl+00}
{Kirkpatrick}, J.~D., et~al. 2000, \aj, 120, 447

\bibitem[\protect\citeauthoryear{{Knapp} et~al.}{{Knapp} et~al.}{2004}]{klf+04}
{Knapp}, G.~R., et~al. 2004, \aj, 127, 3553

\bibitem[\protect\citeauthoryear{{Krishnamurthi}, {Leto}, \&
  {Linsky}}{{Krishnamurthi} et~al.}{1999}]{kll99}
{Krishnamurthi}, A., {Leto}, G.,  \& {Linsky}, J.~L. 1999, \aj, 118, 1369

\bibitem[\protect\citeauthoryear{{Leggett} et~al.}{{Leggett}
  et~al.}{2002}]{lgf+02}
{Leggett}, S.~K., et~al. 2002, \apj, 564, 452

\bibitem[\protect\citeauthoryear{{Leto} et~al.}{{Leto} et~al.}{2000}]{lpl+00}
{Leto}, G., {Pagano}, I., {Linsky}, J.~L., {Rodon{\`o}}, M.,  \& {Umana}, G.
  2000, \aap, 359, 1035

\bibitem[\protect\citeauthoryear{{Liebert} et~al.}{{Liebert}
  et~al.}{2003}]{lkc+03}
{Liebert}, J., {Kirkpatrick}, J.~D., {Cruz}, K.~L., {Reid}, I.~N., {Burgasser},
  A., {Tinney}, C.~G.,  \& {Gizis}, J.~E. 2003, \aj, 125, 343

\bibitem[\protect\citeauthoryear{{Lodieu}, {Scholz}, \& {McCaughrean}}{{Lodieu}
  et~al.}{2002}]{lsm02}
{Lodieu}, N., {Scholz}, R.-D.,  \& {McCaughrean}, M.~J. 2002, \aap, 389, L20

\bibitem[\protect\citeauthoryear{{Lodieu} et~al.}{{Lodieu}
  et~al.}{2005}]{lsm+05}
{Lodieu}, N., {Scholz}, R.-D., {McCaughrean}, M.~J., {Ibata}, R., {Irwin}, M.,
  \& {Zinnecker}, H. 2005, \aap, 440, 1061

\bibitem[\protect\citeauthoryear{{Mart{\'{\i}}n} \& {Ardila}}{{Mart{\'{\i}}n}
  \& {Ardila}}{2001}]{ma01}
{Mart{\'{\i}}n}, E.~L.,  \& {Ardila}, D.~R. 2001, \aj, 121, 2758

\bibitem[\protect\citeauthoryear{{Mart{\'{\i}}n} et~al.}{{Mart{\'{\i}}n}
  et~al.}{1999}]{mdb+99}
{Mart{\'{\i}}n}, E.~L., {Delfosse}, X., {Basri}, G., {Goldman}, B.,
  {Forveille}, T.,  \& {Zapatero Osorio}, M.~R. 1999, \aj, 118, 2466

\bibitem[\protect\citeauthoryear{{Mart{\'{\i}}n} et~al.}{{Mart{\'{\i}}n}
  et~al.}{2001}]{mdm+01}
{Mart{\'{\i}}n}, E.~L., {Dougados}, C., {Magnier}, E., {M{\'e}nard}, F.,
  {Magazz{\`u}}, A., {Cuillandre}, J.-C.,  \& {Delfosse}, X. 2001, \apjl, 561,
  L195

\bibitem[\protect\citeauthoryear{{McCaughrean} et~al.}{{McCaughrean}
  et~al.}{2004}]{mcs+04}
{McCaughrean}, M.~J., {Close}, L.~M., {Scholz}, R.-D., {Lenzen}, R., {Biller},
  B., {Brandner}, W., {Hartung}, M.,  \& {Lodieu}, N. 2004, \aap, 413, 1029

\bibitem[\protect\citeauthoryear{{Mohanty} \& {Basri}}{{Mohanty} \&
  {Basri}}{2003}]{mb03}
{Mohanty}, S.,  \& {Basri}, G. 2003, \apj, 583, 451

\bibitem[\protect\citeauthoryear{{Mohanty} et~al.}{{Mohanty}
  et~al.}{2002}]{mbs+02}
{Mohanty}, S., {Basri}, G., {Shu}, F., {Allard}, F.,  \& {Chabrier}, G. 2002,
  \apj, 571, 469

\bibitem[\protect\citeauthoryear{{Mullan} \& {MacDonald}}{{Mullan} \&
  {MacDonald}}{2001}]{mm01}
{Mullan}, D.~J.,  \& {MacDonald}, J. 2001, \apj, 559, 353

\bibitem[\protect\citeauthoryear{{Nakajima} et~al.}{{Nakajima}
  et~al.}{1995}]{nok+95}
{Nakajima}, T., {Oppenheimer}, B.~R., {Kulkarni}, S.~R., {Golimowski}, D.~A.,
  {Matthews}, K.,  \& {Durrance}, S.~T. 1995, \nat, 378, 463

\bibitem[\protect\citeauthoryear{{Nakajima}, {Tsuji}, \&
  {Yanagisawa}}{{Nakajima} et~al.}{2004}]{nty04}
{Nakajima}, T., {Tsuji}, T.,  \& {Yanagisawa}, K. 2004, \apj, 607, 499

\bibitem[\protect\citeauthoryear{{Neuh{\"a}user} et~al.}{{Neuh{\"a}user}
  et~al.}{1999}]{nbc+99}
{Neuh{\"a}user}, R., et~al. 1999, \aap, 343, 883

\bibitem[\protect\citeauthoryear{{Neupert}}{{Neupert}}{1968}]{neu68}
{Neupert}, W.~M. 1968, \apjl, 153, L59

\bibitem[\protect\citeauthoryear{{Oppenheimer} et~al.}{{Oppenheimer}
  et~al.}{1995}]{okm+95}
{Oppenheimer}, B.~R., {Kulkarni}, S.~R., {Matthews}, K.,  \& {Nakajima}, T.
  1995, Science, 270, 1478

\bibitem[\protect\citeauthoryear{{Osten} et~al.}{{Osten} et~al.}{2006}]{ohb+06}
{Osten}, R.~A., {Hawley}, S.~L., {Bastian}, T.~S.,  \& {Reid}, I.~N. 2006,
  \apj, 637, 518

\bibitem[\protect\citeauthoryear{{Pizzolato} et~al.}{{Pizzolato}
  et~al.}{2003}]{pmm+03}
{Pizzolato}, N., {Maggio}, A., {Micela}, G., {Sciortino}, S.,  \& {Ventura}, P.
  2003, \aap, 397, 147

\bibitem[\protect\citeauthoryear{{Raedler} et~al.}{{Raedler}
  et~al.}{1990}]{rwm+90}
{Raedler}, K.-H., {Wiedemann}, E., {Meinel}, R., {Brandenburg}, A.,  \&
  {Tuominen}, I. 1990, \aap, 239, 413

\bibitem[\protect\citeauthoryear{{Reid} et~al.}{{Reid} et~al.}{2003a}]{rca+03}
{Reid}, I.~N., et~al. 2003a, \aj, 126, 3007

\bibitem[\protect\citeauthoryear{{Reid} et~al.}{{Reid} et~al.}{2003b}]{rcl+03}
{Reid}, I.~N., et~al. 2003b, \aj, 125, 354

\bibitem[\protect\citeauthoryear{{Reid} et~al.}{{Reid} et~al.}{2000}]{rkg+00}
{Reid}, I.~N., {Kirkpatrick}, J.~D., {Gizis}, J.~E., {Dahn}, C.~C., {Monet},
  D.~G., {Williams}, R.~J., {Liebert}, J.,  \& {Burgasser}, A.~J. 2000, \aj,
  119, 369

\bibitem[\protect\citeauthoryear{{Reid} et~al.}{{Reid} et~al.}{1999}]{rkg+99}
{Reid}, I.~N., {Kirkpatrick}, J.~D., {Gizis}, J.~E.,  \& {Liebert}, J. 1999,
  \apjl, 527, L105

\bibitem[\protect\citeauthoryear{{Reid} et~al.}{{Reid} et~al.}{2002}]{rkl+02}
{Reid}, I.~N., {Kirkpatrick}, J.~D., {Liebert}, J., {Gizis}, J.~E., {Dahn},
  C.~C.,  \& {Monet}, D.~G. 2002, \aj, 124, 519

\bibitem[\protect\citeauthoryear{{Reid}}{{Reid}}{2003}]{rei03}
{Reid}, N. 2003, \mnras, 342, 837

\bibitem[\protect\citeauthoryear{{Reiners} \& {Basri}}{{Reiners} \&
  {Basri}}{2006}]{rb06}
{Reiners}, A.,  \& {Basri}, G. 2006, ArXiv Astrophysics e-prints

\bibitem[\protect\citeauthoryear{{Rosner}, {Golub}, \& {Vaiana}}{{Rosner}
  et~al.}{1985}]{rgv85}
{Rosner}, R., {Golub}, L.,  \& {Vaiana}, G.~S. 1985, \araa, 23, 413

\bibitem[\protect\citeauthoryear{{Ruiz}, {Leggett}, \& {Allard}}{{Ruiz}
  et~al.}{1997}]{rla97}
{Ruiz}, M.~T., {Leggett}, S.~K.,  \& {Allard}, F. 1997, \apjl, 491, L107

\bibitem[\protect\citeauthoryear{{Rutledge} et~al.}{{Rutledge}
  et~al.}{2000}]{rbm+00}
{Rutledge}, R.~E., {Basri}, G., {Mart{\'i}n}, E.~L.,  \& {Bildsten}, L. 2000,
  ApJ, 538, L141

\bibitem[\protect\citeauthoryear{{Saar} \& {Linsky}}{{Saar} \&
  {Linsky}}{1985}]{sl85}
{Saar}, S.~H.,  \& {Linsky}, J.~L. 1985, \apjl, 299, L47

\bibitem[\protect\citeauthoryear{{Salim} \& {Gould}}{{Salim} \&
  {Gould}}{2003}]{sg03}
{Salim}, S.,  \& {Gould}, A. 2003, \apj, 582, 1011

\bibitem[\protect\citeauthoryear{{Salim} et~al.}{{Salim} et~al.}{2003}]{slr+03}
{Salim}, S., {L{\'e}pine}, S., {Rich}, R.~M.,  \& {Shara}, M.~M. 2003, \apjl,
  586, L149

\bibitem[\protect\citeauthoryear{{Scholz} \& {Jayawardhana}}{{Scholz} \&
  {Jayawardhana}}{2006}]{sj06}
{Scholz}, A.,  \& {Jayawardhana}, R. 2006, \apj, 638, 1056

\bibitem[\protect\citeauthoryear{{Scholz} et~al.}{{Scholz}
  et~al.}{2003}]{sml+03}
{Scholz}, R.-D., {McCaughrean}, M.~J., {Lodieu}, N.,  \& {Kuhlbrodt}, B. 2003,
  \aap, 398, L29

\bibitem[\protect\citeauthoryear{{Schweitzer} et~al.}{{Schweitzer}
  et~al.}{2001}]{sgh+01}
{Schweitzer}, A., {Gizis}, J.~E., {Hauschildt}, P.~H., {Allard}, F.,  \&
  {Reid}, I.~N. 2001, \apj, 555, 368

\bibitem[\protect\citeauthoryear{{Smith} et~al.}{{Smith} et~al.}{2003}]{sth+03}
{Smith}, V.~V., et~al. 2003, \apjl, 599, L107

\bibitem[\protect\citeauthoryear{{Tinney}}{{Tinney}}{1998}]{tin98}
{Tinney}, C.~G. 1998, \mnras, 296, L42

\bibitem[\protect\citeauthoryear{{Tinney}, {Burgasser}, \&
  {Kirkpatrick}}{{Tinney} et~al.}{2003}]{tbk03}
{Tinney}, C.~G., {Burgasser}, A.~J.,  \& {Kirkpatrick}, J.~D. 2003, \aj, 126,
  975

\bibitem[\protect\citeauthoryear{{Tinney}, {Delfosse}, \& {Forveille}}{{Tinney}
  et~al.}{1997}]{tdf97}
{Tinney}, C.~G., {Delfosse}, X.,  \& {Forveille}, T. 1997, \apjl, 490, L95

\bibitem[\protect\citeauthoryear{{Tsvetanov} et~al.}{{Tsvetanov}
  et~al.}{2000}]{tgz+00}
{Tsvetanov}, Z.~I., et~al. 2000, \apjl, 531, L61

\bibitem[\protect\citeauthoryear{{Vrba} et~al.}{{Vrba} et~al.}{2004}]{vhl+04}
{Vrba}, F.~J., et~al. 2004, \aj, 127, 2948

\bibitem[\protect\citeauthoryear{{West} et~al.}{{West} et~al.}{2004}]{whw+04}
{West}, A.~A., et~al. 2004, \aj, 128, 426

\bibitem[\protect\citeauthoryear{{Wilking}, {Greene}, \& {Meyer}}{{Wilking}
  et~al.}{1999}]{wgm99}
{Wilking}, B.~A., {Greene}, T.~P.,  \& {Meyer}, M.~R. 1999, \aj, 117, 469

\bibitem[\protect\citeauthoryear{{Zapatero Osorio} et~al.}{{Zapatero Osorio}
  et~al.}{2000}]{zbm+00}
{Zapatero Osorio}, M.~R., {B{\'e}jar}, V.~J.~S., {Mart{\'{\i}}n}, E.~L.,
  {Rebolo}, R., {Barrado y Navascu{\'e}s}, D., {Bailer-Jones}, C.~A.~L.,  \&
  {Mundt}, R. 2000, Science, 290, 103

\end{thebibliography}

\clearpage
\begin{deluxetable}{llcrrrccccccl}
\tablecolumns{13}
\tabcolsep0.05in\footnotesize
\tablecaption{Radio Observations of M,L,T Dwarfs
\label{tab:obs}}
\tablehead {
\colhead {RA}                   &
\colhead {Dec}                  &
\colhead {SpT}             &
\colhead {$J$}                  &
\colhead {$K$}                  &
\colhead {$\pi$}                &
\colhead {$F_{\rm \nu,R}$}      &
\colhead {$v{\rm sin}i$}        &
\colhead {$L_{\rm bol}$}        &
\colhead {$L_{\rm H\alpha}/L_{\rm bol}$}    &
\colhead {$R$}                  &       
\colhead {$T$}                  &
\colhead {Notes}                \\
\colhead {}                     &
\colhead {}                     &
\colhead {}                     &
\colhead {(mag)}                &
\colhead {(mag)}                &
\colhead {(mas)}                &
\colhead {($\mu$Jy)}            &
\colhead {(km s$^{-1}$)}        &
\colhead {($L_\odot$)}          &
\colhead {}                     &
\colhead {($R_\odot$)}          &
\colhead {(K)}        
}
\startdata
%
%
\ra{16}{26}{19.2} & \dec{-24}{24}{16} & M5   & 16.35 & 11.73 & 6.25   & $<150$          & \nod & $-1.32$ & \nod      & \nod  & \nod  & [1]           \\
\ra{00}{24}{44.2} & \dec{-27}{08}{24} & M5.5 &  9.26 &  8.23 & 135.3  & $161\pm 15$     & \nod & \nod    & \nod      & \nod  & \nod  & LHS 1070ABC   \\
\ra{04}{35}{16.1} & \dec{-16}{06}{57} & M7   & 10.40 &  9.34 & 116.3  & $<48$	        & \nod & \nod    & $-4.71^a$ & \nod  & \nod  & LP 775-31     \\
\ra{04}{40}{23.2} & \dec{-05}{30}{08} & M7   & 10.68 &  9.56 & 102.0  & $<39$	        & \nod & \nod    & $-4.28^a$ & \nod  & \nod  & LP 655-48     \\
\ra{07}{52}{23.9} & \dec{+16}{12}{16} & M7   & 10.83 &  9.82 & 95.2   & $<39$	        & \nod & \nod    & \nod      & \nod  & \nod  & LP 423-31     \\
\ra{14}{56}{38.3} & \dec{-28}{09}{47} & M7   &  9.96 &  8.92 & 152.4  & $270\pm 40$     & 8    & $-3.29$ & $-5.22$   & \nod  & 2600  & [2], LHS 3003 \\ 
\ra{16}{55}{35.3} & \dec{-08}{23}{40} & M7   &  9.78 &  8.83 & 154.5  & $<24$	        & 9    & $-3.21$ & $-4.06$   & 0.113 & 2707  & [3], VB8      \\
\ra{01}{48}{38.6} & \dec{-30}{24}{40} & M7.5 & 12.28 & 11.24 & 54.3   & $<45$	        & \nod & \nod    & \nod      & \nod  & \nod  &               \\
\ra{03}{31}{30.2} & \dec{-30}{42}{38} & M7.5 & 11.37 & 10.28 & 82.6   & $<72$	        & \nod & \nod    & $-4.21^a$ & \nod  & \nod  & LP 888-18     \\
\ra{04}{17}{37.5} & \dec{-08}{00}{01} & M7.5 & 12.17 & 11.05 & 57.5   & $<36$	        & \nod & \nod    & \nod      & \nod  & \nod  &               \\
\ra{04}{29}{18.4} & \dec{-31}{23}{57} & M7.5 & 10.89 &  9.80 & 103.1  & $<48$	        & \nod & \nod    & \nod      & \nod  & \nod  &               \\
\ra{15}{21}{01.0} & \dec{+50}{53}{23} & M7.5 & 12.00 & 10.92 & 62.1   & $<39$	        & \nod & \nod    & \nod      & \nod  & \nod  &               \\
\ra{10}{16}{34.7} & \dec{+27}{51}{50} & M7.5 & 11.95 & 10.95 & 63.3   & $<45$	        & 7    & \nod    & $-4.70$   & \nod  & \nod  & [1], LHS 2243 \\
\ra{19}{16}{57.6} & \dec{+05}{09}{02} & M8   &  9.95 &  8.81 & 174.2  & $<81$	        & 6.5  & $-3.35$ & $-4.32$   & \nod  & 2700  & [3], VB10     \\
\ra{15}{34}{57.0} & \dec{-14}{18}{48} & M8   & 11.39 & 10.31 & 90.9   & $<111$	        & \nod & $-3.39$ & $-4.80^a$ & \nod  & 2500  & [2]           \\
\ra{10}{48}{14.2} & \dec{-39}{56}{09} & M8   &  9.55 &  8.45 & 247.5  & $140\pm	40$     & 25   & $-3.39$ & $-4.92^a$ & \nod  & 2500  & [2]           \\
                  &                   &      &       &       &        & $(29.6\pm 1)\times 10^4$ &   &   &           &       &       & flare         \\
\ra{11}{39}{51.1} & \dec{-31}{59}{21} & M8   & 12.67 & 11.49 & 50     & $<99$	        & \nod & $-3.39$ & $-4.22^a$ & \nod  & 2500  & [2]           \\
\ra{18}{43}{22.1} & \dec{+40}{40}{21} & M8   & 11.30 & 10.27 & 70.9   & $<48$	        & \nod & $-3.10$ & \nod      & \nod  & \nod  & LHS 3406      \\
\ra{00}{19}{26.3} & \dec{+46}{14}{08} & M8   & 12.61 & 11.47 & 51.3   & $<33$	        & \nod & \nod    & \nod      & \nod  & \nod  &               \\ 
\ra{03}{50}{57.4} & \dec{+18}{18}{07} & M8   & 12.95 & 11.76 & 43.7   & $<105$	        & 4    & \nod    & $-4.06$   & \nod  & 2550  & LP 413-53     \\
\ra{04}{36}{10.4} & \dec{+22}{59}{56} & M8   & 13.76 & 12.19 & 7.14   & $<45$	        & 8    & \nod    & $-4.36^a$ & \nod  & \nod  & CFHT-BD-tau-2 \\
\ra{05}{17}{37.7} & \dec{-33}{49}{03} & M8   & 12.00 & 10.82 & 68.0   & $<54$	        & \nod & \nod    & \nod      & \nod  & \nod  & DENIS $0517-33$ \\
\ra{20}{37}{07.1} & \dec{-11}{37}{57} & M8   & 12.28 & 11.26 & 59.5   & $<33$	        & \nod & \nod    & \nod      & \nod  & \nod  &               \\
\ra{15}{01}{08.3} & \dec{+22}{50}{02} & M8.5 & 11.80 & 10.74 & 94.4   & $190\pm 15$     & 60   & $-3.59$ & $-5.03$   & 0.097 & 2319  & [1], TVLM 513-46 \\
                  &                   &      &       &       &        & $980\pm 40$     &      &         &           &       &       & flare         \\
\ra{03}{35}{02.1} & \dec{+23}{42}{36} & M8.5 & 12.26 & 11.26 & 52.1   & $<69$	        & 30   & \nod    & $-4.63$   & \nod  & 2475  &               \\
\ra{18}{35}{37.9} & \dec{+32}{59}{55} & M8.5 & 10.27 &  9.15 & 176.4  & $525\pm 15$     & \nod & \nod    & \nod      & \nod  & \nod  & LSR 1835+3259 \\
\ra{14}{54}{28.0} & \dec{+16}{06}{05} & M8.5 & 11.14 & 10.02 & 101.9  & $<30$	        & \nod & $-3.39$ & \nod      & \nod  & 2440  & [3], GJ 569Ba \\
\ra{14}{54}{28.0} & \dec{+16}{06}{05} & M9   & 11.65 & 10.43 & 101.9  & $<30$	        & \nod & $-3.56$ & \nod      & \nod  & 2305  & [3], GJ 569Bb \\
\ra{08}{53}{36.2} & \dec{-03}{29}{32} & M9   & 11.18 &  9.97 & 117.6  & $<81$	        & 12   & $-3.49$ & $-4.30$   & 0.101 & 2441  & [1,3], LHS 2065 \\
\ra{01}{09}{51.2} & \dec{-03}{43}{26} & M9   & 11.70 & 10.42 & 90.9   & $<33$	        & \nod & \nod    & $-4.52^a$ & \nod  & \nod  & LP 647-13     \\
\ra{04}{34}{15.2} & \dec{+22}{50}{31} & M9   & 13.74 & 11.87 & 7.14   & $<69$	        & 7    & \nod    & $-4.55^a$ & \nod  & \nod  & CFHT-BD-tau-1 \\
\ra{04}{36}{38.9} & \dec{+22}{58}{12} & M9   & 13.70 & 12.34 & 7.14   & $<57$	        & 12   & \nod    & $-3.79^a$ & \nod  & \nod  & CFHT-BD-tau-3 \\
\ra{05}{37}{25.9} & \dec{-02}{34}{32} & M9   & 18.22 & 17.00 & 2.84   & $<66$	        & \nod & $-3.08$ & $-3.00$   & \nod  & 2460  & SOri 55       \\
\ra{17}{07}{23.4} & \dec{-05}{58}{24} & M9   & 12.06 & 10.71 & \nod   & $<48$	        & \nod & \nod    & $-5.82^a$ & \nod  & \nod  &               \\
\ra{00}{19}{45.8} & \dec{+52}{13}{18} & M9   & 12.82 & 11.62 & 53.5   & $<42$	        & \nod & \nod    & \nod      & \nod  & \nod  &               \\ 
\ra{03}{39}{35.2} & \dec{-35}{25}{44} & M9   & 10.75 &  9.52 & 201.4  & $74\pm 13$      & 28   & $-3.79$ & $-5.26$   & 0.093 & 2138  & [4], LP 944-20 \\
                  &                   &      &       &       &        & $2600\pm 200$   &      &         &           &       &       & flare         \\
\ra{04}{43}{37.6} & \dec{+00}{02}{05} & M9.5 & 12.52 & 11.17 & 65.4   & $<60$	        & \nod & \nod    & \nod      & \nod  & \nod  &               \\ 
\ra{00}{24}{24.6} & \dec{-01}{58}{20} & M9.5 & 11.86 & 10.58 & 86.6   & $83\pm 18$      & 34   & $-3.45$ & $<-6.01$  & 0.103 & 2495  & [1], BRI 0021-0214 \\
\ra{00}{27}{42.0} & \dec{+05}{03}{41} & M9.5 & 16.08 & 14.87 & 13.8   & $<75$	        & 13   & $-3.62$ & $-3.39$   & 0.097 & 2302  & [1], PC 0025+0447 \\
\ra{03}{45}{43.1} & \dec{+25}{40}{23} & L0   & 13.92 & 12.67 & 37.1   & $<87$	        & 25   & $-3.56$ & \nod      & 0.098 & 2364  & [1]           \\
\ra{07}{46}{42.5} & \dec{+20}{00}{32} & L0.5 & 11.78 & 10.47 & 81.9   & $<48$	        & 24   & $-3.62$ & $-5.24$   & 0.097 & 2302  &               \\
\ra{06}{02}{30.4} & \dec{+39}{10}{59} & L1   & 12.30 & 10.86 & 94.3   & $<30$	        & \nod & \nod    & \nod      & \nod  & \nod  & LSR 0602+3910 \\
\ra{13}{00}{42.5} & \dec{+19}{12}{35} & L1   & 12.72 & 11.62 & 71.9   & $<87$	        & \nod & \nod    & \nod      & \nod  & \nod  &               \\
\ra{18}{07}{15.9} & \dec{+50}{15}{31} & L1.5 & 12.93 & 11.60 & 68.5   & $<39$	        & \nod & \nod    & \nod      & \nod  & \nod  &               \\
\ra{02}{13}{28.8} & \dec{+44}{44}{45} & L1.5 & 13.49 & 12.21 & 53.5   & $<30$	        & \nod & \nod    & \nod      & \nod  & \nod  &               \\
\ra{20}{57}{54.0} & \dec{-02}{52}{30} & L1.5 & 13.12 & 11.72 & 61.7   & $<36$	        & \nod & \nod    & \nod      & \nod  & \nod  & DENIS $2057-02$ \\
\ra{04}{45}{53.8} & \dec{-30}{48}{20} & L2   & 13.41 & 11.98 & 60.2   & $<66$	        & \nod & \nod    & \nod      & \nod  & \nod  &               \\
\ra{01}{09}{01.5} & \dec{-51}{00}{49} & L2   & 12.23 & 11.09 & 100    & $<111$	        & \nod & $-3.89$ & \nod      & \nod  & 2100  & [2]           \\
\ra{13}{05}{40.1} & \dec{-25}{41}{10} & L2   & 13.41 & 11.75 & 53.6   & $<27$           & 60   & $-3.57$ & $-5.23$   & 0.098 & 2354  & [3], Kelu 1   \\
\ra{05}{23}{38.2} & \dec{-14}{03}{02} & L2.5 & 13.08 & 11.64 & 74.6   & $<39$           & \nod & \nod    & \nod      & \nod  & \nod  &               \\ 
\ra{05}{23}{38.2} & \dec{-14}{03}{02} & L2.5 & 13.08 & 11.64 & 74.6   & $231\pm 14$     & \nod & \nod    & \nod      & \nod  & \nod  &               \\
\ra{17}{21}{03.9} & \dec{+33}{44}{16} & L3   & 13.62 & 12.49 & 65.8   & $<48$	        & \nod & \nod    & \nod      & \nod  & \nod  &               \\
\ra{02}{51}{14.9} & \dec{-03}{52}{45} & L3   & 13.06 & 11.66 & 82.6   & $<36$           & \nod & \nod    & \nod      & \nod  & \nod  &               \\
\ra{21}{04}{14.9} & \dec{-10}{37}{36} & L3   & 13.84 & 12.37 & 58.1   & $<24$           & \nod & \nod    & \nod      & \nod  & \nod  &               \\ 
\ra{00}{45}{21.4} & \dec{+16}{34}{44} & L3.5 & 13.06 & 11.37 & 96.2   & $<39$           & \nod & \nod    & \nod      & \nod  & \nod  &               \\
\ra{00}{36}{16.1} & \dec{+18}{21}{10} & L3.5 & 12.47 & 11.06 & 114.2  & $134\pm 16$     & 15   & $-3.93$ & 0         & 0.091 & 1993  & [1,5]         \\
                  &                   &      &       &       &        & $720\pm 40$     &      &         &           &	     &       & flare         \\ 
\ra{17}{05}{48.3} & \dec{-05}{16}{46} & L4   & 13.31 & 12.03 & 93.4   & $<45$	        & \nod & \nod    & \nod      & \nod  & \nod  & DENIS $1705-05$ \\
\ra{14}{24}{39.0} & \dec{+09}{17}{10} & L4   & 15.69 & 14.17 & 31.7   & $<96$	        & 17.5 & $-4.04$ & $-5.07$   & 0.090 & 1885  & [1], GD 165B  \\
\tablebreak
\ra{06}{52}{30.7} & \dec{+47}{10}{34} & L4.5 & 13.54 & 11.69 & 90.1   & $<33$           & \nod & \nod    & \nod      & \nod  & \nod  &               \\ 
\ra{22}{24}{43.8} & \dec{-01}{58}{52} & L4.5 & 14.07 & 12.02 & 88.1   & $<33$           & \nod & $-4.13$ & \nod      & 0.089 & 1792  &               \\ 
\ra{01}{41}{03.2} & \dec{+18}{04}{50} & L4.5 & 13.88 & 12.49 & 79.4   & $<30$           & \nod & \nod    & \nod      & \nod  & \nod  &               \\ 
\ra{08}{35}{42.5} & \dec{-08}{19}{23} & L5   & 13.17 & 11.14 & 109.9  & $<30$	        & \nod & $-4.09$ & \nod      & \nod  & 1700  &               \\
\ra{01}{44}{35.3} & \dec{-07}{16}{14} & L5   & 14.19 & 12.27 & 74.6   & $<33$           & \nod & \nod    & \nod      & \nod  & \nod  &               \\ 	
\ra{02}{05}{03.4} & \dec{+12}{51}{42} & L5   & 15.68 & 13.67 & 37.0   & $<48$           & \nod & \nod    & \nod      & \nod  & \nod  &               \\
\ra{00}{04}{34.8} & \dec{-40}{44}{05} & L5   & 13.11 & 11.40 & 104.7  & $<45$           & 32.5 & $-4.00$ & $-5.23$   & 0.090 & 1923  & LHS 102B      \\ 
\ra{15}{07}{47.6} & \dec{-16}{27}{38} & L5   & 12.82 & 11.31 & 136.4  & $<57$	        & \nod & $-4.23$ & \nod      & 0.088 & 1703  &               \\
\ra{12}{28}{15.2} & \dec{-15}{47}{34} & L5   & 14.38 & 12.77 & 49.4   & $<87$	        & 22   & $-4.19$ & $-5.75$   & 0.089 & 1734  & [1], DENIS $1228-15$ \\
\ra{15}{15}{00.8} & \dec{+48}{47}{41} & L6   & 14.06 & 12.56 & 108.7  & $<27$	        & \nod & \nod    & \nod      & \nod  & \nod  &               \\
\ra{04}{39}{01.0} & \dec{-23}{53}{08} & L6.5 & 14.41 & 12.82 & 92.6   & $<42$	        & \nod & \nod    & \nod      & \nod  & \nod  &               \\
\ra{02}{05}{29.4} & \dec{-11}{59}{29} & L7   & 14.59 & 13.00 & 50.6   & $<30$           & 22   & $-4.34$ & $<-6.16$  & 0.088 & 1601  & DENIS $0205-11$ \\
\ra{00}{30}{30.0} & \dec{-14}{50}{33} & L7   & 16.28 & 14.48 & 37.4   & $<57$ 	        & \nod & \nod    & \nod      & \nod  & \nod  &               \\
\ra{17}{28}{11.4} & \dec{+39}{48}{59} & L7   & 15.99 & 13.91 & 41.5   & $<54$	        & \nod & \nod    & \nod      & \nod  & \nod  &               \\
\ra{08}{25}{19.6} & \dec{+21}{15}{52} & L7.5 & 15.10 & 13.03 & 93.8   & $<45$	        & \nod & $-4.61$ & \nod      & 0.088 & 1372  &               \\
\ra{04}{23}{48.5} & \dec{-04}{14}{03} & L7.5 & 14.46 & 12.93 & 65.9   & $<42$	        & \nod & \nod    & \nod      & \nod  & \nod  &               \\
\ra{22}{52}{10.7} & \dec{-17}{30}{13} & L7.5 & 14.31 & 12.90 & 120.5  & $<30$	        & \nod & \nod    & \nod      & \nod  & \nod  & DENIS $2252-17$ \\
\ra{09}{29}{33.6} & \dec{+34}{29}{52} & L8   & 16.60 & 14.64 & 45.5   & $<42$	        & \nod & \nod    & \nod      & \nod  & \nod  &               \\
\ra{15}{23}{22.6} & \dec{+30}{14}{56} & L8   & 16.06 & 14.35 & 53.7   & $<45$	        & \nod & $-4.60$ & \nod      & 0.088 & 1376  & Gl 584C       \\ 
\ra{16}{32}{29.1} & \dec{+19}{04}{41} & L8   & 15.87 & 14.00 & 65.6   & $<54$	        & 30   & $-4.65$ & $-6.23$   & 0.088 & 1335  &               \\
\ra{01}{51}{41.5} & \dec{+12}{44}{30} & T0.5 & 16.57 & 15.18 & 46.7   & $<51$           & \nod & $-4.68$ & \nod      & \nod  & 1300  &               \\
\ra{22}{04}{10.5} & \dec{-56}{46}{57} & T1   & 12.29 & 11.35 & 275.8  & $<79$           & 28   & $-4.71$ &           & 0.091 & 1276  & [6], $\epsilon$Ind Ba \\
\ra{02}{07}{42.8} & \dec{+00}{00}{56} & T4.5 & 16.80 & 15.41 & 34.8   & $<39$	        & \nod & $-4.82$ & \nod      & \nod  & 1200  &               \\
\ra{05}{59}{19.1} & \dec{-14}{04}{48} & T4.5 & 13.80 & 13.58 & 97.7   & $<27$	        & \nod & $-4.53$ & \nod      & \nod  & 1425  &               \\
\ra{15}{34}{49.8} & \dec{-29}{52}{27} & T5.5 & 14.90 & 14.84 & 73.6   & $<63$           & \nod & $-5.00$ & \nod      & \nod  & 1070  &               \\
\ra{16}{24}{14.4} & \dec{+00}{29}{16} & T6   & 15.49 & 15.52 & 90.9   & $<36$           & \nod & $-5.16$ & \nod      & \nod  & 975   &               \\
\ra{22}{04}{10.5} & \dec{-56}{46}{57} & T1   & 13.23 & 13.53 & 275.8  & $<79$           & \nod & $-5.35$ &           & 0.096 & 854   & [6], $\epsilon$Ind Bb \\
\ra{13}{46}{46.4} & \dec{-00}{31}{50} & T6.5 & 16.00 & 15.77 & 68.3   & $<105$	        & \nod & $-5.00$ & \nod      & \nod  & 1075  & [1]           \\
\ra{10}{47}{53.9} & \dec{+21}{24}{23} & T6.5 & 15.82 & 16.41 & 94.7   & $<45$	        & \nod & $-5.35$ & \nod      & \nod  & 900   &               \\
\ra{06}{10}{35.1} & \dec{-21}{51}{17} & T7   & 14.20 & 14.30 & 173.2  & $<69$           & \nod & $-5.21$ & \nod      & \nod  & 950   & [3], Gl 229B  \\
\ra{12}{17}{11.1} & \dec{-03}{11}{13} & T7.5 & 15.86 & 15.89 & 90.8   & $<111$	        & \nod & $-5.32$ & \nod      & \nod  & 900   &               \\
\ra{04}{15}{19.5} & \dec{-09}{35}{06} & T8   & 15.70 & 15.43 & 174.3  & $<45$           & \nod & $-5.73$ & \nod      & \nod  & 700   & 
\enddata
\tablecomments{Properties of the M,L,T dwarfs presented in this paper.
The columns are (left to right): (i) right acsension, (ii)
declination, (iii) spectral type, (iv) $J$-band magnitude, (v)
$K$-band magnitude, (vi) parallax, (vii) radio flux, (viii) rotation
velocity, (ix) bolometric luminosity, (x) H$\alpha$ activity, (xi) 
radius, (xii) surface temperature,and (xiii) notes: [1] \citet{ber02}, 
[2] \citet{bp05}, [3] \citet{kll99}, [4] \citet{bbb+01}, [5] 
\citet{brr+05}, [6] \citet{abb+05}.  $^a$: $L_{\rm H\alpha}/L_{\rm 
bol}$ calculated from the H$\alpha$ line equivalent width using the 
conversion of \citet{whw+04}.  References for source properties:
 \citet{wgm99}, 
\citet{cr02}, \citet{crl+03}, \citet{lsm+05}, \citet{rca+03}, 
\citet{dhv+02}, \citet{mb03}, \citet{rei03}, \citet{giz02},
\citet{fs04}, \citet{dh01}, \citet{gmr+00}, \citet{mdm+01},
\citet{lgf+02}. \citet{glm+04}, \citet{gkl+02}, \citet{rkl+02},
\citet{rcl+03}, \citet{cmj+05}, \citet{sg03}, \citet{lkc+03},
\citet{zbm+00}, \citet{ma01}, \citet{nbc+99}, \citet{gmg+92},
\citet{klf+04}, \citet{hck+02}, \citet{mdb+99}, \citet{gr06},
\citet{bdk+04}, \citet{slr+03}, \citet{kir05}, \citet{lsm02},
\citet{bai04}, \citet{rla97}, \citet{tin98}, \citet{kdm+04},
\citet{sgh+01}, \citet{rkg+00}, \citet{krl+00}, \citet{khl93},
\citet{kma+03}, \citet{brl+05}, \citet{dtf+99}, \citet{tdf97},
\citet{kdm+01}, \citet{sml+03}, \citet{mcs+04}, \citet{tbk03},
\citet{sth+03}, \citet{bwk+00}, \citet{bkm+03}, \citet{bkl+03},
\citet{bkr+03}, \citet{tgz+00}, \citet{bbm+03}, \citet{bkb+99},
\citet{nok+95}, \citet{okm+95}.}
\end{deluxetable}

\clearpage
\begin{deluxetable}{lccccccc}
\tablecolumns{8}
\tabcolsep0.12in\footnotesize
\tablecaption{Physical Properties of M and L dwarfs
\label{tab:phys}}
\tablehead {
\colhead {}       &
\colhead {}       &
\multicolumn{3}{c}{Quiescent} &
\multicolumn{3}{c}{Flaring}   \\\cline{3-5}\cline{6-8}
\colhead {Object}       &
\colhead {Sp.~Type}     &
\colhead {$B$}          &
\colhead {$R$}          &
\colhead {$n_e$}        &
\colhead {$B$}          &
\colhead {$R$}          &
\colhead {$n_e$}        \\
\colhead {}             &
\colhead {}             &
\colhead {(G)}          &
\colhead {(cm)}         &
\colhead {(cm$^{-3}$)}  &
\colhead {(G)}          &
\colhead {(cm)}         &
\colhead {(cm$^{-3}$)}  
}
\startdata
LHS 1070     & M5.5 & $<135$ & $<3.0\times 10^9$   & $>3.3\times 10^9$    & \nod           & \nod                & \nod              \\
DENIS 1048   & M8   & \nod   & \nod                & \nod                 & $2\times 10^3$ & $<3\times 10^8$     & $3\times 10^{11}$ \\
TVLM 513     & M8.5 & $<35$  & $<3.0\times 10^9$   & $>1.5\times 10^{10}$ & $630$          & $1.9\times 10^{10}$ & $1.5\times 10^5$  \\
LSR 1835+32  & M8.5 & $<13$  & $<1.7\times 10^9$   & $>5.8\times 10^9$    & \nod           & \nod                & \nod              \\
LP 944-20    & M9.5 & $<95$  & $<1.2\times 10^9$   & $>1.3\times 10^9$    & $135$          & $7.1\times 10^9$    & $6.9\times 10^7$  \\
BRI 0021     & M9.5 & $95$   & $3.0\times 10^9$    & $5.3\times 10^8$     & \nod           & \nod                & \nod              \\
2M $0523-14$ & L2.5 & $55$   & $4.6\times 10^9$    & $2.2\times 10^9$     & \nod           & \nod                & \nod              \\
2M 0036+18   & L3.5 & $175$  & $1.0\times 10^{10}$ & $1.6\times 10^6$     & $560$          & $1.3\times 10^{10}$ & $3.2\times 10^5$   
\enddata
\tablecomments{Derived magnetic field strengths, as well as electron 
densities and emission region sizes for the M and L dwarfs presented 
in this paper and in the literature.}
\end{deluxetable}

\clearpage

\clearpage

\clearpage

\clearpage

\end{document}